\documentstyle{article}

\newtheorem{theorem}{Theorem}[section]
\newtheorem{define}[theorem]{Definition}
\newenvironment{definition}{\begin{define} \rm}{\end{define}}
\newtheorem{exa}[theorem]{Example}

\newtheorem{lemma}[theorem]{Lemma}
\newtheorem{corollary}[theorem]{Corollary}
\newtheorem{remark}[theorem]{Remark}

\newtheorem{exe}{Exercise}

\def\smallromani{\renewcommand{\theenumi}{\roman{enumi}}
        \renewcommand{\labelenumi}{(\theenumi)}}

\newcommand{\qed}{\hspace*{\fill}\nolinebreak\mbox{$\quad\Box$}}

\newcommand{\rarrow}{\rightarrow}
\newcommand{\rrarrow}{\longrightarrow}
\newcommand{\labtran}[1]{\stackrel{#1}{\rrarrow}}
\newcommand{\slabtran}[1]{\stackrel{#1}{\Longrightarrow}}
\newcommand{\indrule}[2]{\frac{\raisebox{1ex}{$#1$}}{\raisebox{-1.5ex}{$#2$}}} 
 
\newcommand{\la}{\langle}
\newcommand{\ra}{\rangle}
\newcommand{\os}{[\![}
\newcommand{\cs}{]\!]}

\long\def\comment#1{}

\begin{document}

\title{Comparing the Expressive Power of the Synchronous and the Asynchronous 
$\pi$-calculus\thanks{This work has been supported by the HCM project 
``EXPRESS''.}}
\author{Catuscia Palamidessi\\
 DISI, Universit\`a di Genova, via Dodecaneso, 35, 16146 Genova, Italy\\ 
{\tt catuscia@disi.unige.it} }

\maketitle

\footnote{Proceedings of the 24th Annual SIGPLAN-SIGACT Symposium on
  Principles of Programming Languages, Paris, France, January 15--17,
  1997.}

\begin{abstract}
The Asynchronous $\pi$-calculus, as recently proposed by Boudol and,
independently, by Honda and Tokoro, is a subset of the $\pi$-calculus which
contains no explicit operators for choice and output-prefixing.
The communication mechanism of this calculus, however, is powerful enough
to simulate output-prefixing, as shown by Boudol, and input-guarded
choice, as shown recently by Nestmann and Pierce.
A natural question arises, then, whether or not it is possible to embed in it 
the full $\pi$-calculus. We show that this is not possible, 
i.e. there does not exist any uniform, parallel-preserving, translation 
from the $\pi$-calculus into the asynchronous $\pi$-calculus, up to any 
``reasonable'' 
notion of equivalence. This result is based on the incapablity of the
asynchronous $\pi$-calculus of breaking certain symmetries possibly present 
in the initial communication graph. By similar arguments, we prove a 
separation result between the $\pi$-calculus and CCS.
\end{abstract}

\section{Introduction}\label{Intro}

Communication is one of the fundamental ``ingredients''
of concurrent and distributed computation. 
This mechanism can be of two kinds: synchronous and asyn\-chro\-nous. 
The first one is usually understood as {\it simultaneous} 
exchange of information between the two partners; 
an example of it, in ``real life'', is the telephone. 
The latter arises when the action of sending 
a message, and the action of receiving it, 
do not have to occur at the same time. 
An example of it is e.mail. 
Advantages and disadvantages of the one and the other 
method are easy to imagine: more efficient but more expensive the 
first, allowing for more independence the second, etc. 

In the field of models for concurrency, 
it arises naturally the 
question whether these two mechanisms 
are equivalent; i.e., whether they can be 
implemented the one in the other. 
Actually, one direction is clear: 
asynchronous communication can be 
simulated by inserting between each pair of communicating agents
a ``queue''  process 
(see for instance \cite{JJH90}).
The other direction, on the contrary, is not clear and 
researchers in the field seem to have radically different opinions
about it.

The motivation for this work arises from the attempt 
of solving, or at least clarifying, this question. 
The initial guess of the author
was that asynchronous communication is less powerful.
This intuition is 
supported by the example of two people who try 
to take a common decision by using e.mail instead of telephone: 
If they act always in the same way, i.e. send 
at the same time identical mails
and react in the same way to what they read, then
an agreement might never be reached. 
 
Since we were trying to show a separation result, it seemed
convenient to study this problem in the framework of the 
$\pi$-calculus (\cite{MPW92}). 
This is a synchronous paradigm, and a  fragment
of it has been presented recently as  
``asynchronous'' (\cite{Bou92,HT91}). 
We could thus work in a uniform context. But, more important, 
the $\pi$-calculus (and also its asynchronous subset) 
is one of the richest paradigm for concurrency 
introduced so far, hence a separation result
in this context would be more significant. 

The asynchronous $\pi$-calculus differs from the 
$\pi$-cal\-cu\-lus for the lack of the choice and the output prefix
operators. 
The underlying model of interaction among processes, however, 
is the same as in the $\pi$-calculus (handshaking). 
The reason why it is considered asynchronous is 
that, due to the lack of output prefix, an output action 
can only be written ``in parallel'' with other activities, 
thus it is not possible to control when it will actually be 
executed. From the point of view of the process in which 
such an action occurs, it amounts to the impossibility of controlling 
when the message will actually be read by the receiver. 

In recent years the interest in this asynchronous fragment 
has grown, in particular concerning the question of its 
expressiveness. Boudol has shown in \cite{Bou92} that 
the  lazy lambda calculus can still be encoded into it
(as it is the case for the $\pi$-calculus). 
Honda and Tokoro,  and independently Boudol,  have shown 
that output prefix can be simulated
(\cite{HT91,Bou92}). 
Concerning choice, the local (or internal) kind can be easily encoded
(\cite{HT92}). More interestingly, it has been 
proved recently by 
Nestmann and Pierce that also 
input-guarded choice can be encoded (\cite{NP96}). 
Note that this justifies the more recent presentations
of asynchronous $\pi$-calculus, which include 
input-guarded choice as an explicit operator (\cite{BS96,ACS96}).

The only question that remains open is wheth\-er
the asynchronous $\pi$-calculus can simulate
the output-guard\-ed choice (or to be more precise, the 
mixed choice, i.e. the presence of both kinds of guards).
In this work, we show that it is not possible. 
For proving this result, we use techniques from 
the field of Distributed Computing. 
In particular, we show that 
in  symmetric networks, 
it is not possible, with the 
asynchronous $\pi$-calculus, to solve the 
leader election problem, i.e. 
to guarantee that all processes will reach a common agreement 
(elect the leader) in a finite amount of time. 
It is possible, on the contrary, to solve this problem with the 
full $\pi$-calculus. 

The use of this technique has been inspired by the work of 
Boug\'e (\cite{Bo88}), 
who has shown a similar separation result 
concerning the $CSP$ (\cite{Ho78}) and the fragment of 
$CSP$ with no output guards, $CSP_{\it in}$.
The main difference is that the 
asynchronous $\pi$-calculus is a much richer language 
than $CSP_{\it in}$, hence our result is not a consequence 
of the result of Boug\'e. 
Some evidence of this is provided by the fact 
that a second result of Boug\'e, 
concerning the non-encodability of $CSP_{\it in}$
into its choice-free fragment, does not extend 
to the context of the $\pi$-calculus, as shown 
by  the above mentioned result of Nestmann and Pierce.
For a more extended and technical discussion 
about the relation with \cite{Bo88}
see the last section of this paper.
  
Another problem we consider is
the question  to what extent the  $\pi$-calculus
is more powerful than its ``ancestor'' CCS (\cite{Mi89}). 
Also CCS can be seen as a subset of the $\pi$-calculus;
the main difference is the presence, in the latter,  of a mechanism 
of name passing, which allows to change dynamically the 
structure of the communication graph. 
By similar arguments as above
(existence/non-existence of symmetric electoral systems)
we show that this capability makes the  $\pi$-cal\-cu\-lus
strictly more expressive than CCS. 

The rest of the paper is organized as follows:  
next section recalls basic definitions. 
Section 3 reformulates in the setting of 
the $\pi$-calculus the notions of symmetric and electoral system. 
Section 4 shows the main result of the paper, i.e. 
the non-existence of symmetric electoral systems in 
the asynchronous $\pi$-calculus. Section 5 discusses 
existence of symmetric electoral systems for
the synchronous case, i.e. the $\pi$-calculus
and CCS.
Section 6 interprets previous results as 
non-encodability results. 
Section 7 discusses related work and concludes.

\section{Preliminaries}
In this section we recall the definition of the $\pi$-calculus, 
the asynchronous $\pi$-calculus, 
and the notion of hypergraph, which will be used to 
represent the communication structure of a network of processes. 

\subsection{The $\pi$-calculus}
Many variants of the $\pi$-calculus have been proposed. 
Here we basically follow the presentation given in 
\cite{BS96,San95}. The main difference with the 
original version (\cite{MPW92}) is the absence of the matching operator, 
and a construct for guarded choice instead of free choice.

Let  ${\cal N}$ be  a countable set of {\it names}, 
$x, y,\ldots$. The  set of prefixes,   
$\alpha, \beta,\ldots$,
and the set of $\pi$-calculus processes,   
$P,Q,\ldots$, are defined by the following 
syntax: 

\[
\begin{array}{rlcl}
{\it Prefixes}&\alpha&\mbox{::=}&x(y)\;\;|\;\; \bar{x}y \;\;|\;\; \tau \\
{\it Processes}&P&\mbox{::=}& \sum_i\alpha_i.P_i \;\;|\;\; 
    \nu x P \;\;|\;\; P|P \;\;|\;\; !P
\end{array}
\]

Prefixes represent the basic actions of processes: 
$x(y)$ is the {\it input} of the (formal) name $y$ 
from channel $x$; 
$\bar{x}y$ is the {\it output} of the name $y$ 
on channel $x$; 
$\tau$ stands for any silent (non-communication) action. 

The process $\sum_i\alpha_i.P_i$ represents guarded (global) choice   
and it is usually assumed to be finite. 
We will use the abbreviations 
${\bf 0}$ ({\it inaction}) to represent the empty sum, 
$\alpha.P$ ({\it prefix}) to represent sum on one element only, and 
$P+Q$ for the binary sum. 
The symbols $\nu x$, $|$, and $!$ are the {\it restriction}, 
the {\it parallel}, and the {\it replication} operator, respectively. 

The operators $\nu x$ and  $y(x)$ are $x$-{\it binders}, 
i.e. in the processes  $\nu x P$ and $y(x).P$ the occurrences 
of $x$ in $P$ are considered {\it bounded}, 
with the usual rules of scoping. 
The {\it free names} of $P$, i.e. those names which do 
not occur in the scope of any binder, 
are denoted by  ${\it fn}(P)$. 
The {\it alpha-conversion} of bounded names is defined as usual, 
and the renaming (or substitution) $P\{y/x\}$ is defined as the 
result of replacing all occurrences of $x$ in $P$ by $y$, possibly 
applying alpha-conversion to avoid capture. 

The operational semantics is specified via a transition system 
labeled by {\it actions} $\mu, \mu'\ldots$. 
These are given by the following grammar: 
\[
\begin{array}{rlcl}
{\it Actions} &\mu &\mbox{::=}& x(y) \;\;|\;\; \bar{x}y \;\;|\;\; \bar{x}(y)
   \;\;|\;\; \tau
\end{array}
\]
Essentially, we have all the actions corresponding to prefixes, 
plus the {\it bounded output} $\bar{x}(y)$. This is introduced to model 
{\it scope extrusion}, i.e. the result of sending to another process 
a private ($\nu$-bounded) name. 
The bounded names of an action $\mu$, ${\it bn}(\mu)$,
are defined as follows:
${\it bn}(x(y))={\it bn}(\bar{x}(y))=\{y\}$; 
${\it bn}(\bar{x}y)={\it bn}(\tau)=\emptyset$.
Furthermore, we will indicate by $n(\mu)$ all the 
{\it names} which occur in $\mu$.

In literature there have been considered two 
definitions for the transition system of the $\pi$-calculus, 
which induce two different semantics: the {\it early} and the 
{\it late} bisimulation semantics. 
Here we choose to present the first one because the 
early bisimulation is coarser than the other, 
but it should be noted that the results 
of this paper are independent from the
bisimulation semantics adopted at this point.
(No notion of bisimulation can identify an electoral system and a 
non-electoral one.)

The rules for the early semantics 
are given in Table~\ref{TS}. 
The symbol $\equiv$ used in Rule {\sc Cong}
stands for {\it structural congruence}, 
a form of equivalence which identifies 
``statically'' two processes. 
Again, there are several definition of this
relation in literature. For our purposes
we do not need a very rich notion, we 
will just use it to simplify the presentation. 
Hence we only assume this congruence to satisfy the following: 
\begin{enumerate}
\smallromani
\item $P\equiv Q$ if $Q$ can be 
obtained from $P$ by alpha-renaming, notation $P\equiv_\alpha Q$,
\item $P|Q\equiv Q|P$,
\item $(P|Q)|R\equiv P|(Q|R)$,
\item $(\nu x P) | Q\equiv \nu x (P|Q)$ if $x\not \in {\it fv}(Q)$.
\end{enumerate}

\begin{table} 
\begin{center}
\begin{tabular}{|@{\ \ \ }ll@{\ \ }|}  
\hline
\mbox{}&\mbox{}
\\
{\sc I-Sum}
  &$\sum_i\alpha_i.P_i \labtran{x(z)} P_j\{z/y\}$
    \ \ \ \ $\alpha_j=x(y)$
\\
&\mbox{}
\\
{\sc O/$\tau$-Sum}
  &$\sum_i\alpha_i.P_i \labtran{\alpha_j} P_j$
    \ \ \ \ $\alpha_j=\bar{x}y$ or $\alpha_j=\tau$
\\
&\mbox{}
\\
{\sc Open}
 &$\indrule{P\labtran{\bar{x}y}P'}{\nu y P\labtran{\bar{x}(y)} P'}$
  \ \ \ \ $x\neq y$
\\
&\mbox{}
\\
{\sc Res}
 &$\indrule{P\labtran{\mu}P'}{\nu y P\labtran{\mu}\nu y P'}$
  \ \ \ \ $y\not\in n(\mu)$
\\
&\mbox{}
\\
{\sc Par}
 &$\indrule{P\labtran{\mu}P'}{P | Q\labtran{\mu}  P' | Q}$
  \ \ \ \ ${\it bn}(\mu)\cap{\it fn}(Q) =\emptyset$
\\
&\mbox{}
\\
{\sc Com}
 &$\indrule{P\labtran{x(y)}P'\ \ \ \ Q\labtran{\bar{x}y}Q'}
 {P | Q\labtran{\tau} P' | Q'}$
\\
&\mbox{}
\\
{\sc Close}
 &$\indrule{P\labtran{x(y)}P'\ \ \ \ Q\labtran{\bar{x}(y)}Q'}
  {P | Q\labtran{\tau} \nu y(P' | Q')}$
\\
&\mbox{}
\\
{\sc Rep}
 &$\indrule{P|!P\labtran{\mu}P'}
  {!P \labtran{\mu} P' }$
\\
&\mbox{}
\\
{\sc Cong}
 &$\indrule{P\equiv P' \ \ \ \ P'\labtran{\mu}Q' \ \ \ \ Q'\equiv Q}
  {P\labtran{\mu}Q}$
\\
&\mbox{} 
\\
\hline
\end{tabular}
\caption{The early-instantiation transition system of the $\pi$-calculus.}
\label{TS}
\end{center}
\end{table}

\subsection{The asynchronous $\pi$-calculus}

In accordance with \cite{HT91,Bou92}, we consider the following definition of
the asynchronous $\pi$-calculus 
($\pi_a$-calculus for short).

\[
{\it Processes}\;\;\; P\;\; \mbox{::=}\;\; \bar{x}y \;\;|\;\; x(y).P \;\;|\;\; 
    \nu x P \;\;|\;\; P|P \;\;|\;\; !P
\]

The difference wrt 
the $\pi$-calculus is that $\sum_i\alpha_i.P_i$
is replaced by the output-action process $\bar{x}y$ and by the 
input-prefix process $x(y).P$. The rule for the output-action process
is described in Table~\ref{TSA}, where $\bf 0$ stands again 
for inaction (see \cite{Bou92} for the encoding of inaction 
into the $\pi_a$-calculus.) All the rules for the 
other operators are like in Table~\ref{TS}.

\begin{table} 
\begin{center}
\begin{tabular}{|llll|}  
\hline
\mbox{   }&&&\mbox{   }
\\
&{\sc Out}
  &$\bar{x}y \labtran{\bar{x}y}{\bf 0}$&\mbox{   }
\\
&&&\mbox{   }
\\
\hline
\end{tabular}
\caption{The output rule for the $\pi_a$-calculus.}
\label{TSA}
\end{center}
\end{table}

Note that the $\pi_a$-calculus is a proper subset of the 
$\pi$-calculus. The output-action process $\bar{x}y$, 
in fact, could be equivalently replaced 
by the  special case of output prefix $\bar{x}y.{\bf 0}$.

\subsection{Hypergraphs and automorphisms}
In this section we recall the 
definition of {\it hypergraph}, which generalize the concept of graph 
essentially by allowing an arc to connect more than 
two nodes. 

A hypergraph is a pair $H=\la N, X, t \ra$
where $N,X$ are finite sets whose elements are called 
{\it nodes} and ({\it hyper}){\it arcs} respectively, 
and $t$ ({\it type}) is a function which assigns to each $x\in X$ 
a set of nodes, representing the nodes {\it connected} by 
$x$. We will also use the notation $x:n_1,\ldots,n_k$ 
to indicate $t(x)= \{n_1,\ldots,n_k\}$. 

The concept of graph automorphism 
extends naturally to hypergraphs:
Given a hypergraph $H=\la N, X, t \ra$, 
an {\it automorphism} on $H$ is a pair $\sigma=\la \sigma_N,\sigma_X\ra$ 
such that $\sigma_N:N\rarrow N$ and $\sigma_X:X\rarrow X$ are 
permutations which preserve the type of arcs, 
namely for each $x\in X$, if $x: n_1,\ldots,n_k$, 
then $\sigma_X(x):\sigma_N(n_1),\ldots,\sigma_N(n_k).$

It is easy to see that the {\it composition} of automorphisms, 
defined componentwise  as $\sigma\circ\sigma'= \la 
\sigma_N\circ\sigma'_N,\sigma_X\circ\sigma'_X\ra$,
is still an automorphism. Its identity is 
the pair of identity functions on $N$ and $X$, i.e.
${\it id}=\la {\it id}_N,{\it id}_X\ra$.
It is easy to show that the set of automorphisms
on $H$ with the composition forms a group.

Given $H$ and $\sigma$ as above, the {\it orbit} 
of $n\in N$ generated by $\sigma$ is defined as the set of nodes 
in which the various iterations of $\sigma$ map $n$, 
namely: 
\[
{\it O}_\sigma(n) = \{ n, \sigma(n), \sigma^2(n),\ldots , \sigma^{h-1}(n)\}
\]
\noindent
where $\sigma^i$ represents the composition of $\sigma$ with itself 
$i$ times, and $\sigma^h={\it id}$.
It is possible to show that the orbits generated by $\sigma$ constitute 
a partition of $N$.

\section{Electoral and Symmetric systems}

In this section we adapt to the $\pi$-calculus 
(a simplified version of)
the notions of electoral system and symmetric network as given 
by Boug\'e in \cite{Bo88}.

\subsection{Election of a leader in a network}
We first need to define the concepts of {\it network computation}
and its {\it projection} over a component of the network. 
A network is a system of parallel process $P=P_1|P_2|\ldots|P_k$.
A computation $C$ for this system is a  (possibly $\omega$-infinite)
sequence of transitions\footnote{For the sake of keeping the 
notation simple, we assume that each binder $\nu x$
generated by a possible application of the {\sc Close} rule, 
is pushed ``to the top level'' by repeated applications of 
the properties $(i)$ and $(iv)$ of $\equiv$.  
Furthermore, we do not represent explicitly 
the binders at the top level; we just assume that 
the network will never perform a visible action on 
one of the names restricted by those binders.}
\[
\begin{array}{lcl}
P_1|P_2|\ldots|P_k 
 &\labtran{\mu^{0}}
   &P^{1}_1 | P^{1}_2|\ldots|P^{1}_k
   \\
 &\labtran{\mu^{1}} 
   &P^{2}_1 | P^{2}_2|\ldots|P^{2}_k
   \\
 &\vdots
   \\
 &\labtran{\mu^{n-1}} 
   &P^{n}_1 | P^{n}_2|\ldots|P^{n}_k
   \\
 & (\;\labtran{\mu^{n}}\;\;
   &\ldots\;) 
\end{array}
\]
\noindent
with $n\geq 0$. 
We will represent it  
also by $C: P\slabtran{\tilde\mu}P^{n}$
(by $C: P\slabtran{\tilde\mu}$ if it is infinite),
$\tilde\mu$ being the sequence 
$\mu^{0}\mu^{1}\ldots\mu^{n-1} (\mu^{n}\ldots)$, and 
$P^n$ being the process $P^{n}_1 | P^{n}_2|\ldots|P^{n}_k$.
The relation $C\preceq C'$ ($C'$ {\it extends} $C$) is defined as usual. 
Namely, let $C: P\slabtran{\tilde\mu}P^{n}$. Then $C\preceq C'$
iff there exists $C'':P^{n}\slabtran{\tilde\mu'}P^{n+n'}$ 
or $C'':P^{n}\slabtran{\tilde\mu'}$, and
$C'= CC''$ (identifying the two occurrences of $P^n$). 
We will denote by $C'\setminus C$ the continuation $C''$.
The notation $C\prec C'$ will indicate that $C'$ is a {\it strict 
extension} of $C$.
Note that if $C$ is infinite then it cannot be strictly 
extended, because we
admit only $\omega$-infinite (i.e. not transfinite) 
computations.

Given $P$ and $C$ as above, the projection 
of $C$ over $P_i$, 
${\it Proj}(C,P_i)$\footnote{For the sake of brevity 
here we have introduced an abuse of notation: the projection is not a 
function of $C$, but of the sequence of proof-trees which generate $C$.}
is defined as the ``contribution'' of $P_i$ 
to the computation. 
More formally, ${\it Proj}(C,P_i)$
is the computation 
\[
P_i\;\;
\slabtran{\tilde\mu^{0}}\;\;
P^{1}_i\;\;
\slabtran{\tilde\mu^{1}}\;\;
P^{2}_i\;\;
\slabtran{\tilde\mu^{2}}\;\;
\ldots\;\;
\slabtran{\tilde\mu^{n-1}}\;\;
P^{n}_i \;\;
(\slabtran{\tilde\mu^{n}}\ldots) 
\]
\noindent 
where, depending on the application of the 
rule ({\sc Par}, {\sc Com}, or {\sc Close})
which generate the $m+1$-th transition of $C$,
$P^{m}_i
\slabtran{\tilde\mu^{m}}
P^{m+1}_i$ 
is:
\begin{itemize}
\item 
$P^{m}_i
\labtran{\mu^m}
P^{m+1}_i$, 
if the rule is {\sc Par} with this transition as premise,
\item 
$P^{m}_i
\labtran{\alpha}
P^{m+1}_i$, 
if the rule is {\sc Com} or {\sc Close} and this transition is
one of the two premises.
\item empty (and therefore $P^{m}_i=P^{m+1}_i$ and $\tilde\mu^{m}$ is 
emp\-ty)
if, in the $m+1$-th transition of $C$,
$P^{m}_i$ is idle, i.e. it does not appear in the 
premises of the rule. 
\end{itemize}

To give the definition of electoral system, 
we assume the existence of a special output channel name, 
{\it o}, shared by all processes. Furthermore we assume 
that $\cal N$ contains the natural numbers, 
which will represent the identifier of processes 
in a network. 

Intuitively, an electoral system has the 
property that at each possible run
the processes will agree sooner or later on 
``which of them has to be the leader'', 
and will communicate this decision to the 
``external world'' by using the channel {\it o}. 

\begin{definition}\label{electoral}{\bf (Electoral system)}
A process $P=P_1|P_2|\ldots|P_k$
is an electoral system if 
for every computation $C$ for $P$ 
there exists an extension $C'$ of $C$ 
and there exists $n\in\{1,\ldots,k\}$
(the ``leader'')
such that for each $i\in \{1,\ldots,k\}$
the projection ${\it Proj}(C',i)$
contains one output action 
of the form $\bar{\it o}n$, and 
no extension of $C'$ contain
any other action of the form $\bar{\it o}m$, 
with $m\neq n$.
\end{definition}

Note that for such a system 
an infinite computation $C$ must contain already
all the output actions of each process
because $C$ cannot be strictly extended.  

\subsection{Symmetric networks}

In order to define the notion of {\it symmetric network},
we have to consider its initial communication structure, 
which we will represent as an hypergraph.
Intuitively, the nodes represent the 
processes, and the arcs the free communication channels, 
connecting the nodes which share them. It will be
convenient, although not necessary, 
not to consider as an arc the``channel to the external world'' $o$. 
 
\begin{definition}{\bf (Hypergraph associated to a network)}
Given a network 
$P=P_1|P_2|\ldots|P_k$, 
the hypergraph associated to $P$ 
is $H(P)=\la N,X,t\ra$ with $N=\{1,\ldots,k\}$, 
$X={\it fn}(P)\setminus\{o\}$, and 
for each $x\in X$, $t(x)= \{ n| x\in{\it fn}(P_n)\}$.
\end{definition}

Intuitively, a system $P$ is symmetric with respect to an 
automorphism $\sigma$ on $H(P)$ iff  for each $i$
\begin{quote}
{\it the process associated to the node 
$\sigma(i)$ is i\-den\-ti\-cal (modulo alpha-conversion) to the
process obtained by $\sigma$-renaming the process 
associated to the node $i$}.  
\end{quote}
The notion of $\sigma$-renaming is the obvious extension of 
the standard notion of renaming (see the preliminaries). 
More formally, given a process $Q$, first apply alpha-conversion 
so to rename all bounded names into fresh ones, 
extend $\sigma$ to be the identity on these new names, and define 
$\sigma(Q)$ by structural induction as indicated below.
For the sake of simplicity, here we use $\sigma(\cdot)$ to represent both 
$\sigma_N(\cdot)$ and $\sigma_X(\cdot)$. Furthermore we extend 
$\sigma$ on prefixes in the obvious way,  i.e. 
$\sigma(x(y))=\sigma(x)(\sigma(y))$, $\sigma(\bar{x}y)=
\overline{\sigma{\scriptstyle (}x{\scriptstyle )}}\sigma(y)$, 
and $\sigma(\tau)=\tau$.
\[
\begin{array}{rcl}
\sigma(\sum_i\alpha_i.P_i) &=& \sum_i\sigma(\alpha_i).\sigma(P_i)\\
\sigma(\nu x P) &=&\nu x\; \sigma(P)\\
\sigma(P | Q) &=&\sigma(P) | \sigma(Q)\\
\sigma(! P) &=&! \sigma(P)
\end{array}
\]

We are now ready to give the formal definition of symmetric system: 

\begin{definition}{\bf (Symmetric system)}
Consider a  network $P=P_1|P_2|\ldots|P_k$, 
and let $\sigma$ be an isomorphism on its associated hypergraph 
$H(P)=\la N, X,t\ra$. 
We say that $P$ is {\it symmetric wrt $\sigma$}
iff for each node $i\in N$, 
$P_{\sigma(i)}\equiv_\alpha\sigma(P_i)$ holds;
$P$ is {\it symmetric} if it is symmetric  wrt all the automorphisms 
on $H(P)$.
\end{definition}

Note that if $P$ is symmetric wrt $\sigma$ then 
$P$ is symmetric wrt all the powers of $\sigma$.

\section{Symmetric electoral systems: the asynchronous case}

This section contains the main result of the paper, 
which is that, 
for certain communication graphs, 
it is not possible to 
write in $\pi_a$-calculus a symmetric network 
solving the election problem. 

We first need to show that the $\pi_a$-calculus 
enjoyes a certain kind of {\it confluence} property: 

\begin{lemma}\label{confluence}
Let $P$ be a process of the $\pi_a$-calculus. Assume that 
$P$ can make two transitions 
$P\labtran{\mu}Q$ and $P\labtran{\mu'}Q'$,
where $\mu$ is an output action
while $\mu'$ is an input action. 
Then there exists 
$R$ such that $Q\labtran{\mu'}R$ and 
$Q'\labtran{\mu}R$.
\end{lemma}

\noindent
{\bf Proof}
Assume that  $\mu$ is of the form 
$\bar{x}y$ or $\bar{x}(y)$, and that 
$\mu'$ is of the form 
$z(w)$. Observe that $x,y,z$ must be free names in $P$.
The rule which has produced the $\mu$-transition can be only {\sc Out},
{\sc Open}, {\sc Res}, {\sc Par}, {\sc Rep}, or {\sc Cong}. 
In the last (five) cases the 
assumption is again a $\mu$-transition. 
By repeating this reasoning (descending the tree), 
we must arrive to a leaf of the form 
$\bar{x}y\labtran{\bar{x}y}{\bf 0}$. 
Analogously, by descending the 
tree for the $\mu'$-transition we must arrive to 
a leaf of the form 
$z(w).S\labtran{z(w')}S\{w'/w\}$. 
Now, $\bar{x}y$ and $z(w).S$ must be two parallel 
processes in $P$, i.e. there must be a subprocess 
in $P$ of the form  $T[\bar{x}y] | T'[z(w).S]$
(modulo $\equiv$), i.e. 
$P\equiv U[T[\bar{x}y] | T'[z(w).S]]$
(here $T[\;]$, $T'[\;]$ and $U[\;]$ represent  contexts, with the usual 
definition).
Furthermore, the 
$\mu$ and $\mu'$ transitions must have been 
obtained by the application of the rule
{\sc Par} to this subprocess, i.e. 
$Q\equiv U[T[{\bf 0}]| T'[z(w).S]]$ and 
$Q'\equiv U[T[\bar{x}y] | T'[S\{w'/w\}]]$.
By applying again the rule {\sc Par} 
(plus all the other rules in the trees for the 
$\mu'$ and the $\mu$ transition respectively) 
we obtain 
the transitions 
$Q\labtran{\mu'} U[T[{\bf 0}] | T'[S\{w'/w\}]]$
and 
$Q'\labtran{\mu} U[T[{\bf 0}] | T'[S\{w'/w\}]]$.
\qed

\bigskip
\noindent

We are now ready to prove the announced non-ex\-is\-tence result.
The intuition is the following: In the attempt to reach 
an agreement about the leader, the processes
of a symmetric network have to ``break the initial symmetry", 
and therefore have to communicate. 
The first such communication, however, can be repeated, 
by the above lemma, and by symmetry, 
by all the pair of processes of the network. 
The result of all these transitions will still lead to a 
symmetric situation.
Thus there is a (infinite) computation in which the processes 
never succeed to break the symmetry, which means no leader 
is elected. 

\begin{theorem}\label{nonexistence api}
Consider a  network $P=P_1|P_2|\ldots|P_k$ in the $\pi_a$-calculus, 
and assume that the associated hypergraph 
$H(P)$ admits an automorphism $\sigma\neq {\it id}$ with only one
orbit, and that $P$ is symmetric wrt $\sigma$.
Then $P$ cannot be an electoral system.
\end{theorem}

\noindent
{\bf Proof}
Assume by contradiction that $P$ is an electoral system. 
We will show that we can then construct an infinite 
increasing sequence
of computations for $P$, 
$C_0\prec C_1\prec\ldots \prec C_h\ldots$, 
such that for each $j$, $C_j:P\slabtran{\tilde\mu^j}P^j$ 
does not contain any  output action 
on {\it o}, and $P^j$is still symmetric wrt 
$\sigma_j$, where $\sigma^j$ is the original 
authomorphism enriched with associations on the 
new names possibly introduced by the communication actions
(for simplicity of notation, in the following $\sigma_j$
will still be indicated as $\sigma$). 
This gives a contradiction, 
because the limit of this sequence is an infinite computation 
for $P$ which does not contain any output action on {\it o}.

We prove the above by induction wrt $h$. In order to understand the 
proof, it is important to notice that
the hypothesis of $\sigma$ generating only one orbit implies that
for each $i\in \{1,2,\ldots,k\}$, 
${\it O}_{\sigma}(i)$ $=$ 
$\{i,\sigma(i),\ldots ,\sigma^{k-1}(i)\}= \{1,2,\ldots,k\}$.

\noindent
{\bf $h=0$)} 
Define $C_0$ to be the empty computation.

\noindent
{\bf $h+1$)}
Given  $C_h:P\slabtran{\tilde\mu^h}P^h$, 
we construct 
$C_{h+1}:P\slabtran{\tilde\mu^{h+1}}P^{h+1}$ as follows. 

Since $P$ is an electoral system, 
it must be possible to extend 
$C_h$ to a computation $C$ which contains ($k$) actions 
$\bar{\it o}n$, for a particular $n\in \{1,\ldots,k\}$.
Observe that the first action $\mu$ of 
$C\setminus C_h$ cannot be $\bar{\it o}n$. 
Otherwise, 
let $P^h_i$ be the component which performs 
this action. Then  $P^h_i$ must contain 
the subprocess $\bar{\it o}n$
and must have no restriction on $n$.
By symmetry, 
$P^h_\sigma(i)\equiv \sigma(P^h_i)$  must contain 
the subprocess $\bar{\it o}\sigma(n)$ and have no restriction 
on $\sigma(n)$. Hence there must be an extension of $C$ 
where the action $\bar{\it o}\sigma(n)$ occurs. 
This implies (for the hypothesis that $P$ is an electoral system),
that $\sigma(n)=n$, and, since $\sigma$ generates only one orbit, that
$\sigma={\it id}$ (and $k=1$). Contradiction.

Hence, $\mu$ must be either $\tau$ or an action on a channel different 
from $o$. Let us distinguishes the two cases.
\begin{description}

\item[$\mu\neq \tau$)]
Let $P^h_i$ be the component which performs 
this action. Let $P^{h+1}_i$ be such that
\[
P^h_i\;\;\labtran{\mu}\;\;P^{h+1}_i 
\]
\noindent
By symmetry we also have
\[
\begin{array}{rcl}
P^h_{\sigma(i)}&\labtran{\sigma(\mu)}&P^{h+1}_{\sigma(i)}\\
P^h_{\sigma^2(i)}&\labtran{\sigma^2(\mu)}&P^{h+1}_{\sigma^2(i)}\\
&\vdots&\\
P^h_{\sigma^{k-1}(i)}&\labtran{\sigma^{k-1}(\mu)}&P^{h+1}_{\sigma^{k-1}(i)}
\end{array}
\]
Since $\sigma$ generates only one orbit, 
$P^h\equiv P^h_i\mid P^h_{\sigma(i)}\mid P^h_{\sigma^2(i)}\mid\ldots
\mid P^h_{\sigma^{k-1}(i)}$.
Hence we can compose the displayed transitions into 
a computation
\[
P^h\;\;
\slabtran{\tilde\mu}\;\;
P^{h+1},
\]
\noindent
 where 
$\tilde{\mu}= \mu\sigma(\mu)\sigma^2(\mu)\ldots \sigma^{k-1}(\mu)$
and
$P^{h+1}\equiv P^{h+1}_i\mid P^{h+1}_{\sigma(i)}\mid P^{h+1}_{\sigma^2(i)}\mid\ldots
\mid P^{h+1}_{\sigma^{k-1}(i)}$.
Finally, observe that 
$P^{h+1}$ is still symmetric.

\item[$\mu= \tau$)]
In this case, the transition is the result of a communication 
between two agents. The interesting case is when the 
two agents are in {\it different nodes} of the communication graph. 
(If the agents are inside the same node, say 
$P^h_i$, then we have a transition $P^h_i\;\;\labtran{\tau}\;\;P^{h+1}_i$
and we proceed 
like in previous case.)
Let $P^h_i$ and $P^h_j$ be the two processes, with $i\neq j$. 
We have two transitions 
$P^h_i\labtran{\mu_i}Q_i$ and $P^h_j\labtran{\mu_j}R_j$,
where $\mu_i$ and $\mu_j$ are complementary.
Assume without loss of generality that $\mu_i$ is the input action, and
$\mu_j$ is the output action. 
Since $\sigma$ generates only one orbit, there exists $r\in\{1,\ldots,k-1\}$
such that $j=\sigma^r(i)$. Assume 
for simplicity that $r$ and $k$ are relatively 
prime\footnote{If they are not, then in the rest of the proof 
$k$ has to be replaced by the least $p$ such that $pk=rq$, for some $q$.},
and let $\theta=\sigma^r$. Then  $P^h_{j}=P^h_{\theta(i)}$ and
$R_j=R_{\theta(i)}$. Let us first consider the 
case in which the first step of $C\setminus C_h$
has been produced by an application of the 
{\sc Com} rule. Then we have a transition
\[
P^h_i\mid P^h_{\theta(i)}\;\;\labtran{\tau}\;\;Q_i\mid R_{\theta(i)}
\]
By symmetry, we have that 
$P^h_{\theta(i)}\labtran{\theta(\mu_i)}\theta(Q_i)$.
By Lemma \ref{confluence}
we then have the transitions 
$R_{\theta(i)}\labtran{\theta(\mu_i)}R'$
and 
$\theta(Q_i)\labtran{\mu_j}R'$
for some $R'$. Let us define 
$P^{h+1}_{\theta(i)}=R'$.
By symmetry, we also have 
$P^h_{\theta^2(i)}\equiv P^h_{\theta(j)}\labtran{\theta(\mu_j)}\theta(R_{j})$, 
and $\theta(\mu_i)$, $\theta(\mu_j)$ are complementary, 
hence we can combine them into a transition 
\[
R_{\theta(i)}\mid P^h_{\theta^2(i)}\;\;\labtran{\tau}\;\; P^{h+1}_{\theta(i)}
\mid R_{\theta^2(i)}
\]
\noindent 
with $R_{\theta^2(i)}= \theta(R_{j})$. 
By repeatedly applying this reasoning, we obtain 
\[\begin{array}{rcl}
R_{\theta^2(i)}\mid P^h_{\theta^3(i)}&\labtran{\tau}& 
P^{h+1}_{\theta^2(i)}
\mid R_{\theta^3(i)}\\
&\vdots\\
R_{\theta^{k-2}(i)}\mid P^h_{\theta^{k-1}(i)}&\labtran{\tau}& 
P^{h+1}_{\theta^{k-2}(i)}
\mid R_{\theta^{k-1}(i)}
\end{array}
\]
\noindent 
and 
$R_{\theta^{k-1}(i)}\labtran{\theta^{k-1}(\mu_i)}P^{h+1}_{\theta^{k-1}(i)}$.
Finally, observe that from the transition 
$\theta(Q_i)\labtran{\mu_j}R'$
above we can derive 
$\theta^k(Q_i)\labtran{\theta^{k-1}(\mu_j)}\theta^{k-1}(R')$.
But $\theta^k=\sigma^{kr}={\it id}$, hence we have
$Q_i\labtran{\theta^k(\mu_j)}P^{h+1}_i$, 
where we have defined $P^{h+1}_i$ to be $\theta^{k-1}(R')$.
Therefore we can compose also these transitions, thus ``closing the 
circle'', as we obtain
\[
R_{\theta^{k-1}(i)}\mid Q_i\;\;\labtran{\tau}\;\; 
P^{h+1}_{\theta^{k-1}(i)}
\mid P^{h+1}_i 
\]
The composition of the displayed transitions 
gives us the intended continuation\footnote{Under the assumption that
$r$ and $k$ are relatively 
prime, also $\theta$ has only one orbit. 
If we drop this assumption, and hence we replace 
$k$ by the smallest $p$ such that $pk=rq$ for some $q$, 
then the computation we have constructed involves only the 
processes of the nodes in 
$O_\theta(i)=\{i,\theta(i),\ldots,\theta^{p-1}(i)\}$.
To complete computation we have to repeat the reasoning for 
the other orbits of $\theta$: $O_\theta(\sigma(i))$, 
$O_\theta(\sigma^2(i))$\ldots $O_\theta(\sigma^{q-1}(i))$.}:
\[
\begin{array}{l}
P^h \equiv P^{h}_i|P^{h}_{\theta(i)}|\ldots|P^{h}_{\theta^{k-1}(i)}
\\
\qquad\qquad\qquad\qquad\quad
\slabtran{\tilde{\tau}}
\\
\qquad\qquad\qquad\qquad\qquad
P^{h+1}_i|P^{h+1}_{\theta(i)}|\ldots|P^{h+1}_{\theta^{k-1}(i)}
\end{array}
\]
\noindent
Finally define 
$P^{h+1}=
P^{h+1}_i|P^{h+1}_{\theta(i)}|\ldots|
P^{h+1}_{\theta^{k-1}(i)}$
and observe that it is still symmetric
with respect to $\sigma$. 

Consider now the case in which the 
first step of $C\setminus C_h$ 
is obtained by an application of the {\sc Close}
rule. Then 
the transition would be of the form 
\[
P^h_i\mid P^h_{\theta(i)}\;\;\labtran{\tau}\;\;\nu y(Q_i\mid 
R_{\theta(i)}) 
\]
where $y$ is the name transmitted in the communication.
In order to reason as before we have to eliminate 
the $\nu y$ interposed between $Q_i\mid 
R_{\theta(i)}$ and the rest of the network.
This can be done by applying  $\alpha$-conversion 
and  scope extrusion 
(Rules $(i)$ and $(iv)$ of $\equiv$), 
so to push the restriction operator 
at the top-level of the network. However, by doing this,
we add new (free) names and 
enrich the communication structure of the network.
To preserve the simmetry, we must then dynamically 
enrich $\sigma$ with suitable
associations among these new names, in the obvious way. 
For instance, if a communication action occurs between 
the node $i$ and the node $j$, in which a private name $x$ 
of $i$ is transmitted, then an analogous communication
will happen between the nodes $\sigma(i)$ and $\sigma(j)$, 
with transmission of another private name (of $\sigma(i)$), 
say $y$. Correspondingly, we must add the association 
$\sigma(x) = y$.
\qed
\end{description}

Note  that, for the above result, we could have considered
a simpler (more permissive) notion of electoral system, 
obtained by requiring, in Definition \ref{electoral}, 
that $C'$ contains one (or more) actions of the form $\bar{o}n$, 
instead of requiring it for all the projections of $C'$. 
We have defined the electoral system in that way only to remain
closer to the notion in literature. 

In \cite{Bo88} a more 
permissive notion of symmetry is considered for proving negative results.
Namely, the automorphism $\sigma$ can have more orbits,
provided that they  all have the same cardinality.
An automorphism with this property is called {\it well-balanced}. 
In the framework of \cite{Bo88} this is a significant 
generalization, because the language considered there, 
$CSP_{\it in}$, can have the parallel operator only at the 
top level. Hence the condition of a single orbit, there, 
would impose that {\it all} the parallel processes present 
in the network have the same code (modulo renaming). 

In our framework, on the contrary, we do not have this restriction, 
and the above mentioned generalization is not essential. 
In fact, we can easily extend Theorem~\ref{nonexistence api} to well-balanced 
automorphisms:

\begin{corollary}\label{well balanced}
Consider a  network $P=P_1|P_2|\ldots|P_k$ in the $\pi_a$-calculus, 
and assume that the associated hypergraph 
$H(P)$ admits a well-balanced automorphism $\sigma\neq {\it id}$,
and that $P$ is symmetric wrt $\sigma$.
Then $P$ cannot be an electoral system.
\end{corollary}

\noindent
{\bf Proof}
Assume that $\sigma$ generates $p$ orbits of cardinality $q$, and let 
$i_1,i_2,\ldots,i_p$ be arbitrary nodes from these orbits. 
Consider the processes
\[
\begin{array}{rcl}
Q_1 &=& P_{i_1}| P_{i_2}|\ldots | P_{i_p}\\
Q_2 &=& P_{\sigma(i_1)}| P_{\sigma(i_2)}|\ldots | P_{\sigma(i_p)}\\
&\vdots&\\
Q_{q} &=& P_{\sigma^{q-1}(i_1)}| P_{\sigma^{q-1}(i_2)}|\ldots | 
P_{\sigma^{q-1}(i_p)}
\end{array}
\]
Consider now the network 
$Q=Q_1|Q_2|\ldots|Q_{q}$. Clearly $Q\equiv P$, 
but the associated hypergraph, $H(Q)$, is different. More precisely, 
$H(Q)$ is ``an abstraction'' of  $H(P)$ in the sense that 
certain nodes of  $H(P)$ are ``grouped together'' in the 
same node of  $H(Q)$. (The way this grouping is done 
depends on the choice of $i_1,i_2,\ldots,i_p$ and it
is inessential for this 
proof.) The arcs $X$ of $H(P)$ are the same as the 
ones of $H(Q)$; the type function is the obvious one. 

Now, consider the pair $\theta=\la \sigma_N,\sigma_X\ra$ 
with $\theta_N(1)=2$, $\theta_N(2)=3$,\ldots, $\theta_N(q)=1$,
and  $\theta_X=\sigma_X$. It is easy to see 
that $\theta$ is a well balanced automorphism on $H(Q)$, 
and that $Q$ is symmetric 
wrt $\theta$.
Then apply Theorem~\ref{nonexistence api}, and consider that 
a leader in $P$ determines immediately a leader in $Q$.
\qed

\section{Symmetric electoral systems:\ \ the synchronous case}

In the (synchronous) $\pi$-calculus,
the guarded choice construct makes it possible to 
establish a simultaneous a\-gree\-ment among 
two processes, thus breaking the symmetry. 
The  point is that the  
presence of choice invalidates the confluence 
property of Lemma~\ref{confluence}.

Consider for example the election problem 
in a symmetric network consisting of two nodes $P_0$ and 
$P_1$
only, and two arcs, $x_0$ and $x_1$, connecting them. 
A $\pi$-calculus  specification which solves the problem 
is:
\[
\begin{array}{rcl}
P_i&::&\overline{x_i}(y).\bar{o}i\\
   &  &+\\
   &  &{x_{i\oplus 1}}(y).\bar{o}(i\oplus 1)
\end{array}
\]
\noindent
with $i\in\{1,2\}$ and $\oplus$ being the binary sum. 

The following results shows that with the 
$\pi$-calculus 
the existence of  
symmetric electoral systems is guaranteed 
in a large number of cases: 

\begin{theorem}\label{existence pi}
Let $H$ be a {\it connected} hypergraph 
(i.e. each pair of nodes are connected 
by a sequence of arcs).
Then there exists a symmetric electoral system 
$P$, in the $\pi$ calculus, such that 
$H(P)=H$. 
\end{theorem}

\noindent
{\bf Proof} (Hint) 
One possible algorithm is the following.
Let $k$ be the number of nodes.
The generic process
$P_i$: 
\begin{enumerate}
\item Broadcasts a private name $x_i$ 
to all the other processes (which is
possible thanks to the connectivity hypothesis) and, meanwhile, 
receives the private name $x_j$ of 
each other process $P_j$.
\item Repeats (at most $k$ times) a choice where one guard
is an output action on $x_i$, while the 
others are input actions on the  $x_j$'s.
If at a certain point an 
input is selected, then goes to  4.
\item If this point has been reached, then $P_i$ is the 
leader. It broadcasts this  information to all the other processes, 
outputs $\bar{o}i$ and terminates.
\item Waits to receive the name of the leader. 
Then sends it on $o$ and terminates.
\qed
\end{enumerate}

Note that in the above proof we assume that each process
know what's the total number of processes in the network. 

The mechanisms of name-passing and scope extrusion, which makes it possible 
in the $\pi$-calculus to extend dynamically the communication 
structure of the network, 
are essential for the above result. 
In fact, such result would not 
hold for the ``static subset'' of the $\pi$-calculus
 i.e. CCS~\cite{Mi89},
as shown by the following:

\begin{theorem}\label{nonexistence CCS}
Let $P=P_1|P_2|\ldots |P_k$
be a CCS network 
and let the associated hypergraph $H(P)=\la N, X,t\ra$
admit a
well-balanced automorphism $\sigma$ such that 
$P$ is symmetric wrt $\sigma$ and, for each $n\in N$, 
there exist no $h$ such that $\{n,\sigma^h(n)\}\subseteq t(x)$
for some $x\in X$. 
Then $P$ cannot be an electoral system. 
\end{theorem}
 
\noindent
{\bf Proof} (Hint) 
Let $Q=Q_1|Q_2|\ldots |Q_q$ and $\theta$ be defined as in
Corollary~\ref{well balanced}.
An analysis of the kind of interactions possible between 
$Q_i$ and $Q_\theta^r(i)$ shows that, limited to the 
those transitions, these processes enjoy the confluence property
(Lemma \ref{confluence}). In fact a (parallel) component $P_j$ of 
$Q_i$ can only interact with a (parallel)
component $P_{\sigma^r(h)}$  of $Q_\theta^r(i)$ different 
from the component $P_{\sigma^r(j)}$.
\qed

\section{Uniform encoding} 
In this section we use the above results 
to show the non-encodability of 
the $\pi$-calculus into its asynchronous subsets
and into CCS, under certain requirements on the 
notion of encoding  $\os\cdot\cs$.

There is no agreement on what should be a good notion of encoding, 
ad perhaps indeed there should not be a unique notion, but several, 
depending on the purpose. 
However, it seems reasonable to require at least the two following 
properties:
\begin{enumerate}
\item compositionality,
\item preservation of some intended semantics. 
\end{enumerate}

For a distributed system, however, it seems reasonable to 
strengthen the notion of compositionality on the parallel operator
by requiring 
that it is mapped exactly in the parallel operator, i.e. that 
\begin{eqnarray}\label{parallel pres}
\os P|Q\cs &=& \os P\cs \;|\;\os Q\cs
\end{eqnarray}

Likewise, it seems reasonable to require that the encoding 
``behaves well'' wrt renamings, i.e. 
\begin{eqnarray}\label{renaming pres}
\os \sigma (P) \cs &=& \sigma(\os P\cs)
\end{eqnarray}

\noindent
We will call {\it uniform} an encoding 
which satisfies (\ref{parallel pres}) and (\ref{renaming pres}). 

Concerning the notion of semantics, we call ``reasonable'' 
a semantics which distinguishes two processes $P$ and $Q$ whenever 
in some computation of $P$ the actions on certain intended 
channels are different from those 
of any computation of $Q$. In the following, our intended channel is $o$.

\begin{remark}
There exist no uniform encoding of the $\pi$-calculus 
into the $\pi_a$-calculus preserving a  reasonable semantics.
\end{remark}

\noindent
{\bf Proof}
Uniformity preserves symmetry, and a reasonable semantics 
distinguishes an electoral system from a non-electoral one.
Hence apply Theorems~\ref{existence pi} and \ref{nonexistence api}.
\qed

\begin{remark}
There exist no uniform encoding of the $\pi$-calculus 
into $CCS$  preserving a  reasonable semantics.
\end{remark}

\noindent
{\bf Proof}
Analogous, by Theorems~\ref{existence pi} and \ref{nonexistence CCS}.
\qed

Note that if we relax condition (\ref{parallel pres}), imposing just 
generic compositionality  instead, i.e. 
\begin{eqnarray}\label{comp}
\os P|Q\cs &=&C[\;\os P\cs, \os Q\cs\;]
\end{eqnarray}

\noindent 
with $C[\cdot,\cdot]$ generic context, 
then these non-encodability results do 
not hold anymore. In fact, we could 
give an encoding of the form 
\begin{eqnarray*}
\os P|Q\cs &=&\nu y_1\nu y_2\ldots \nu y_n  (\os P\cs | M | \os Q\cs)
\end{eqnarray*}

\noindent 
where $M$ is a ``monitor'' process  which coordinates the activities of 
$P$ and $Q$, interacting with them via the fresh channels 
$y_1,y_2,\ldots,y_n$. 
The translation of a network $P_1|P_2|\ldots|P_n$ would then be 
a tree with the $P_i$'s as leaves, and the monitors as the other nodes. 
The disadvantage of this solution is that it is not a distributed 
implementation; on the contrary, it is a very centralized one. 

\section{Conclusion and related work}

One way to interpret the results presented in this paper
is that they show that, even in a rich language
like $\pi$-calculus, the full choice cannot 
be implemented into its sublanguage without choice.
Actually, we can easily see that Lemma~\ref{confluence}, 
and therefore Theorem~\ref{nonexistence api},
hold even if we consider a language with both 
input-guarded choice and output-guarded choice, 
but fail when we consider mixed choice 
(input and output guards in the same choice
construct). Hence it is this latter mechanisms
which induces a  separation in expressive power.
This seems to reinforce the impression that 
the mixed choice is a
really difficult mechanism to implement.
So far, the only really distributed, but approximated 
solutions we are aware of 
are the probabilistic methods based on 
randomization (see for instance \cite{FR80}). 

Another way to interpret them is by saying that 
the ``real'', i.e. simultaneous, synchronous communication 
cannot be implemented in the asynchronous one. 
In this sense, the translation of \cite{Bou92} would not be 
acceptable since the randez-vous discipline introduces a delay. 
In this view of things, it is not the choice that is the hard operator: 
mixed choice would be easy to realize if 
real synchronous communication would be available.
It is difficult, however, to argue in favor of this 
interpretation by using the results of this paper, 
because the underlying model of 
the $\pi_a$-calculus formalizes communication via simultaneous 
interaction (i.e. ``handshaking'', via the {\sc Com} rule).
In ongoing work, we are studying the impossibility results 
in the context of a ``real'' model for 
asynchronous communication, like the 
one of Asynchronous ACP (\cite{BKT85}).

The non-existence results of this work 
 hold even if we restrict to {\it fair} computations.
The proof of Theorem~\ref{nonexistence api} in fact 
can be slightly 
modified so that for the construction of $C_{h+1}$ from $C_h$ 
we consider each time a different process in the network. 
In this way, the limit of the sequence is a fair computation.

Our Theorems~\ref{nonexistence api} and \ref{nonexistence CCS}
correspond to Theorems 3.2.1 and 4.2.1 in \cite{Bo88},
for $CSP_{\it in}$ and $CSP$ respectively.
The main difference with those results
is that here we are dealing with much richer  
languages. In particular, both the 
$\pi_a$-calculus and CCS admit the parallel operator inside
every process, and not just at the top-level as it is the case 
for $CSP_{\it in}$ and $CSP$ (at least, for the versions considered in 
\cite{Bo88}: all processes in a network are strictly sequential). 
This leads to an essential difference. Namely,  
the proof of Boug\'e shows that the network can get stucked in the attempt 
to elect a leader: since an output action in $CSP_{\it in}$
can be only sequential, the prefix of a computation which leads to the 
first output action, repeated by all processes, 
brings to a global deadlock.
Our proof, on the contrary, 
shows that the system can run forever 
without reaching an agreement: 
whenever a first output action occurs, 
all the other processes can
execute their corresponding output action as well, 
and so on, thus 
generating an infinite computation which never breaks 
the symmetry.
Another difference is that in the $\pi$-calculus the 
network can evolve dynamically. This is the reason 
why Theorem 4.2.1 in \cite{Bo88} does not hold for 
the $\pi$-calculus (as shown by our Theorem~\ref{existence pi}).
This feature complicates the proof of 
Theorems~\ref{nonexistence api} since we have to take 
into account a corresponding evolution of the 
automorphism. 

The use of the parallel operator as a free 
constructor usually enhances significatively 
the expressive power of 
a language. It is for instance essential for implementing  
choice (at least in a restricted form). 
In fact, Boug\'e has shown in \cite{Bo88} 
that it is not possible to encode 
$CSP_{\it in}$ into $CSP_{no}$
(the sublanguage of $CSP$ with neither input nor output guards
in the choice), 
while Nestmann and Pierce have shown in \cite{NP96} that 
the $\pi_a$-calculus can be embedded into 
its subset with no choice. The crucial point is that 
the parallel operator 
allows to represent the main characteristic of the
choice, namely the simultaneous availability of 
its guards.

\subsection*{Acknowledgements}
I would like to thank 
Ilaria Castellani, Pat Lincoln,
Dale Miller, Uwe Nestmann, 
Prakash Panangaden,
Benjamin Pierce, 
Ro\-sa\-rio Pugliese, 
Scott Smolka and Eugene Stark for stimulating 
and insightful discussions.

\end{document}